# Light emission is fundamentally tied to the quantum coherence of the emitting particle

Aviv Karnieli[1,*], Nicholas Rivera[2], Ady Arie[3] and Ido Kaminer[4]

[1]Raymond and Beverly Sackler School of Physics and Astronomy, Tel Aviv University, Ramat Aviv 69978, Tel Aviv, Israel
[2]Department of Physics, Massachusetts Institute of Technology, Cambridge, Massachusetts 02139, USA
[3]School of Electrical Engineering, Fleischman Faculty of Engineering, Tel Aviv University, Tel Aviv 69978, Israel
[4]Department of Electrical Engineering, Technion–Israel Institute of Technology, Haifa 32000, Israel

*corresponding author: kaminer@technion.ac.il

**Coherent emission of light by free charged particles is ubiquitous in many areas of physics and engineering, with the light's properties believed to be successfully captured by classical electromagnetism in all relevant experimental settings. The advent of interactions between light and free quantum matter waves brought about fundamental questions regarding the role of the particle wavefunction. Here we show that even in seemingly classical experimental regimes, light emission is fundamentally tied to quantum properties of the emitting particles, such as their quantum coherence and correlations. By employing quantum electrodynamics, we unveil the role of the particle's coherent momentum uncertainty, without which decoherence of light becomes dominant. As an example, we consider Cherenkov radiation, envisioned for almost a century as a shockwave of light. We find instead that the shockwave's duration is fundamentally bound from below by the particle's coherent momentum uncertainty due to the underlying entanglement between the particle and light. This quantum optical paradigm opens new capabilities in analytical electron microscopy, enabling the measurement of quantum correlations of shaped electron wavepackets. Our findings also have applications for Cherenkov detectors in particle physics. For example, by measuring spectral photon autocorrelations, one can unveil the particle's wavefunction size, shape and coherence. Such schemes are especially intriguing for many high-energy particles, where other techniques are not available.**

**Introduction**

Excitation of waves by a moving object is ubiquitous in many areas of physics, such as electrodynamics[1], acoustics[2], and hydrodynamics[3] – examples are the Cherenkov effect, sound waves, and ship wakes. These processes are thought to be successfully explained by classical physics, wherein wave interference is often critical for describing the phenomena. For example, in electrodynamics[1,4] radiation emission patterns are predicted by Maxwell's equations.

Such is the case for Cherenkov radiation (CR): the emission of light by free charged particles moving faster than the phase velocity of light in a medium[5]. Since its discovery in 1934, a long-standing hallmark of CR is its manifestation as a 'shockwave' of light[1,6–12], resulting from the coherent temporal interference of radiation at a wide spectral range. Despite the wide applicability of this effect, no experiment has ever directly observed the shockwave dynamics emitted by a single particle. As one of the implications of this work, we shall see that the underlying quantum nature of CR fundamentally limits the shockwave duration in many existing experimental settings, due to the entanglement of the light with the emitting particle.

Looking at the bigger picture in electromagnetism, light emission by free charged particles constitutes a family of effects[10,13] including, for example, transition radiation[14], Smith-Purcell radiation[15], undulator radiation[16] as well as CR. Often called coherent cathodoluminescence (CL), these phenomena are employed in many areas of physics and engineering, from electron microscopes[10,17,18], particle detectors[19,20], free-electron lasers[21,22], engineerable light sources[23–25] and medical imaging[26,27]. As the spectral range of emitted light can be straightforwardly tuned by varying the particle energy, coherent CL is a promising platform for light generation in otherwise inaccessible regimes[21,28,29], such as at THz, UV, and X-ray frequencies.

The broad tunability of coherent CL, alongside recent advances in shaping[30,31], coherent control[32–34] and entanglement[35,36] of free electrons, make it a probe of fundamental light-matter interaction[37–42], and a prominent candidate for quantum measurement[18]. These advancements brought about fundamental questions regarding the role of the particle wavefunction[37–40,43–47] in coherent CL. However, in all relevant experimental settings, coherent CL is still considered as classical[8,48–51] or semi-classical[41,52]. The general expectation is that when the emitting particle is not directly measured[53], the quantum features of its wavefunction[43–47] cannot leave a detectable mark on the emitted light. A milestone of fundamental importance would be, therefore, to identify observables of coherent CL radiation that are both detectable in practical settings and directly depend on the quantum state of the emitting particles. This observation has implications also for general wave phenomena, such as any mechanical waves excited by free moving objects. Can fundamental quantum aspects of a particle affect the patterns of waves in seemingly classical regimes?

Here we introduce the quantum optical paradigm to describe coherent CL and identify the specific measurements that depend on the quantum wave nature of the emitter. By formulating a general quantum theory of spontaneous light emission by charged particles, we show that already in what are generally assumed to be classical regimes, coherent CL can be dominated by quantum features like wavefunction uncertainty, quantum correlations and decoherence. These effects can be exposed in quantum optical measurements such as photon autocorrelations, and even in seemingly classical features such as the emitted pulse duration.

As a surprising implication for the Cherenkov effect, we find that quantum decoherence imposes a fundamental lower bound for the Cherenkov shockwave duration, predicting an uncertainty principle that connects it to the particle momentum uncertainty. Consequently, we identify many practical scenarios in which Cherenkov radiation is not a shockwave.

Our quantum theory of coherent CL has new applications, such as detecting the shape, size and coherence of the emitter's wavefunction by measuring the spectral autocorrelations of the light it emits – thereby gaining information on the wavefunction uncertainty. Our findings can resolve a question which, with the advent of ultrafast electron microscopes, has been frequently asked: what part of the measured energy spread of an electron beam is due to coherent energy uncertainty, and what part is due to incoherent uncertainty. Moreover, our work sheds light on fundamentally new capabilities to measure quantum properties of charged particles which can serve as an alternative to matter wave holography, which is especially important for many high-energy particles observed in Cherenkov detectors, where holographic techniques do not exist. The results presented in this work pave the way towards novel tunable light sources and measurements sensitive to the wavefunction of free charged particles. Our results were first presented in May 2020 in the CLEO2020 conference[54].

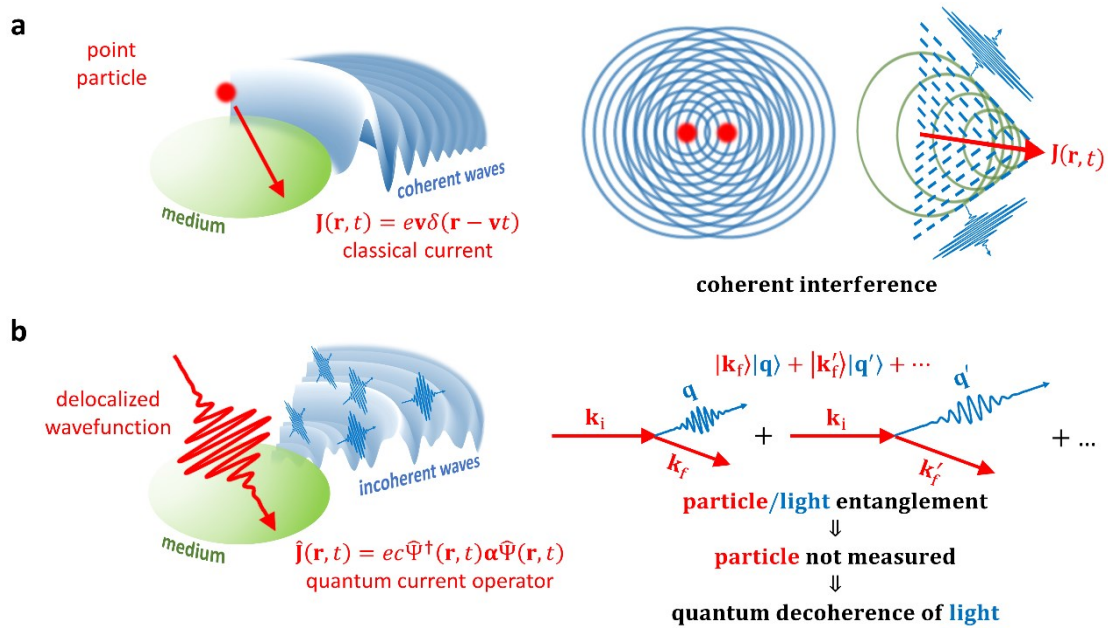

**Fig. 1: Excitation of waves by free particles: classical vs quantum theory.** (a) Classical wave dynamics. A point particle with velocity **v** passes through an optical medium and emits waves that may interfere coherently. The classical emitter current density $\mathbf{J}(\mathbf{r},t) = e\mathbf{v}\delta(\mathbf{r} - \mathbf{v}t)$, emits a temporally coherent shockwave. (b) Quantum description. A quantum particle is described by a delocalized wavefunction $\psi(\mathbf{r},t)$. A current *operator* $\hat{\mathbf{J}}(\mathbf{r},t)$ is then associated with the particle. Even when the initial particle is only described by a single momentum $\mathbf{k}_i$ it may spontaneously emit many wave quanta (momenta **q**, **q'**, …). The waves are then entangled with the particle because of momentum conservation (leaving the final particle having momenta $\mathbf{k}_f$, $\mathbf{k}_f'$, … respectively). When only the emitted waves are observed, this entanglement can lead to quantum decoherence and lack of interference visibility, resulting in the emission of incoherent radiation.

### Results

**Excitation of waves by free particles.** In *classical physics*, waves interfere coherently when they are generated from different point particles constituting an emitter[55], so long as the different emission points are perfectly correlated with each other (Fig. 1a). Particularly, the emission from each individual particle is considered to always be coherent with itself. In *quantum mechanics*, an emitter is described by a spatially varying wavefunction[56]. Following the emission of wave quanta, the particles and waves are in an entangled state, known to cause quantum decoherence[57] (Fig. 1b) if one of the constituents of the bipartite system is not measured. As spontaneous emission of light by free charged particles is usually described classically[8,48–51], it is generally assumed that the abovementioned effect is negligible, on the grounds that the correspondence principle[58] is always valid. This assumption is backed by the small quantum recoil[10] exerted by the photon, amounting to only minor corrections[43–47,59]. It is the purpose of the following analysis to show that under certain common conditions, quantum mechanics fundamentally modifies light emission, even in regimes that are traditionally seen as classical.

Without loss of generality, consider the emitting charged particles to be free electrons. We also consider the emitted electromagnetic field to be in a general optical environment. The initial state is described by a density matrix $\boldsymbol{\rho}_i$, where the electrons have a reduced density matrix $\boldsymbol{\rho}_e$, and the radiation field is found in the vacuum state $|0\rangle$, such that the initial state is separable $\boldsymbol{\rho}_i = \boldsymbol{\rho}_e \otimes |0\rangle\langle 0|$. The interactions between the electrons and the electromagnetic field are governed by the Dirac Hamiltonian: $H_{\text{int}} = ec\boldsymbol{\alpha} \cdot \mathbf{A}$, where $e$ is the electron charge, $c$ the speed of light, $\boldsymbol{\alpha}^i = \gamma^0 \gamma^i$ are the Dirac matrices, and $\mathbf{A}$ is the electromagnetic vector potential operator. Considering a weak coupling between the electrons and photons, the final quantum state of the system, $\boldsymbol{\rho}_f$, is found by first-order time-dependent perturbation theory (see Supplementary Material, section S1).

In general, after the interaction, the electrons are entangled to many photonic modes because emission is allowed for different directions and at many different frequencies. For example, starting from an arbitrary initial wavefunction of a single electron and zero photons, $|\psi_i\rangle = \sum_{\mathbf{k}_i} \varphi_{\mathbf{k}_i} |\mathbf{k}_i\rangle |0\rangle$, and if momentum is conserved – as in CR – the photon can be emitted with different momenta $\mathbf{q} = \mathbf{k}_i - \mathbf{k}_f$, giving an entangled final state

$$|\psi_f\rangle = \sum_{\mathbf{k}_i} \varphi_{\mathbf{k}_i} \sum_{\mathbf{k}_f} M_{\mathbf{k}_i \to \mathbf{k}_f; \mathbf{q}} \, e^{-iE_f t/\hbar} e^{-i\omega_\mathbf{q} t} |\mathbf{k}_f\rangle |\mathbf{q} = \mathbf{k}_i - \mathbf{k}_f\rangle, \qquad (1)$$

where $M_{\mathbf{k}_i \to \mathbf{k}_f; \mathbf{q}}$ is the transition amplitude. Information regarding the electron initial state $\varphi_{\mathbf{k}_i}$ can be extracted by measuring the photon momentum $\mathbf{q} = \mathbf{k}_i - \mathbf{k}_f$ in coincidence with (or post-selection of) an electron momentum $\mathbf{k}_f$. However, this is not the experimental situation of CL: where only the *light* is measured, and the electron degrees of freedom are *traced out*. In this case, both experimental and theoretical evidence suggest that the initial electron wavefunction has no influence on observables of the emitted radiation[37,39,45,60], such as the power spectrum. Below, we will examine this situation carefully, and show how the emitted light autocorrelations *can* be strongly influenced by the single electron wavefunction – even though the power spectrum is not, suggesting a ubiquitous, hidden quantum-ness to the radiation by free electrons.

To describe the photonic final state in the experimental scenario of coherent CL, we calculate the reduced density matrix of the electromagnetic field, $\boldsymbol{\rho}_{\text{ph}} = \text{Tr}_e\{\boldsymbol{\rho}_f\}$, with $\text{Tr}_e$ denoting the partial trace over the electronic state. The electric field autocorrelation is determined by the final photonic state, $\boldsymbol{\rho}_{\text{ph}}$, via the quantum mechanical expectation value $\langle \mathbf{E}^{(-)}(\mathbf{r}',t')\mathbf{E}^{(+)}(\mathbf{r},t)\rangle = \text{Tr}\{\mathbf{E}^{(-)}(\mathbf{r}',t')\mathbf{E}^{(+)}(\mathbf{r},t)\boldsymbol{\rho}_{\text{ph}}\}$, where $\mathbf{E}^{(+)}(\mathbf{r},t)$ and $\mathbf{E}^{(-)}(\mathbf{r},t) = (\mathbf{E}^{(+)}(\mathbf{r},t))^\dagger$ are, respectively, the positive and negative frequency parts of the electric field operator. Instead of the simplified momentum-space picture of Eq. (1), which strictly holds only for CR (see Supplementary Material, Section S8), we employ a more general formalism.

Based on quantum electrodynamical perturbation theory, the formalism holds for *all* coherent CL processes and for an *arbitrary* number of electrons (see Supplementary Material, section S1 for derivation), yielding

$$\langle \mathbf{E}^\dagger(\mathbf{r}',\omega')\mathbf{E}(\mathbf{r},\omega)\rangle = \omega\omega'\mu_0^2 \int d^3\mathbf{R}d^3\mathbf{R}' \mathbf{G}^\dagger(\mathbf{r}',\mathbf{R}',\omega')\mathbf{G}(\mathbf{r},\mathbf{R},\omega)\langle \mathbf{j}^\dagger(\mathbf{R}',\omega')\mathbf{j}(\mathbf{R},\omega)\rangle_e, \quad (2)$$

where $\mathbf{G}(\mathbf{r},\mathbf{r}',\omega)$ is the Dyadic Green's function of Maxwell's equations for the dielectric medium[61,62], and where $\mathbf{E}^{(+)}(\mathbf{r},t) = \int_0^\infty d\omega e^{-i\omega t}\mathbf{E}(\mathbf{r},\omega)$. The quantity $\langle \mathbf{j}^\dagger(\mathbf{r}',\omega')\mathbf{j}(\mathbf{r},\omega)\rangle_e = \mathrm{Tr}\{\boldsymbol{\rho}_e \mathbf{j}^\dagger \mathbf{j}\}$ is the expectation value, with respect to the *emitter initial state*, of the correlations in the current density operator $\mathbf{j}(\mathbf{r},t) = ec\Psi^\dagger \boldsymbol{\alpha}\Psi$, where $\Psi(\mathbf{r},t)$ is the emitter spinor field operator described in second quantization. From here onwards, we assume that the particles propagate as wavepackets with a well-defined carrier velocity $\mathbf{v}_0$ (the paraxial approximation, where the particle dispersion is linearized).

Now, let us constrain the discussion to the seemingly classical regime, where photon recoils $\hbar q$ are much smaller than electron momenta $p_e$. This constraint is applicable to a vast number of effects, including all cases in which the emitter is relativistic, all free-electron nanophotonic light sources, and all free-electron sources in the microwave and radio frequency ranges. In general, this derivation applies to both the single- and many-particle emitter states, described via second quantization of the emitter. The current correlations in Eq. 2 can then be written as (see Supplementary Material section S2 for derivation)

$$\langle \mathbf{j}(\mathbf{x}')\mathbf{j}(\mathbf{x})\rangle = e^2 \mathbf{v}_0\mathbf{v}_0 \left[G_e^{(2)}(\mathbf{x}',\mathbf{x}) + \delta(\mathbf{x}-\mathbf{x}')G_e^{(1)}(\mathbf{x},\mathbf{x})\right], \quad (3)$$

where $\mathbf{x} = \mathbf{r} - \mathbf{v}_0 t$ and $\mathbf{x}' = \mathbf{r}' - \mathbf{v}_0 t'$. In Eq. (3), we define the first and second-order correlation functions of the emitter $G_e^{(1)}(\mathbf{x}',\mathbf{x}) = \sum_\sigma \mathrm{Tr}\{\boldsymbol{\rho}_e \psi_\sigma^\dagger(\mathbf{x}')\psi_\sigma(\mathbf{x})\}$ and $G_e^{(2)}(\mathbf{x}',\mathbf{x}) = \sum_{\sigma'}\sum_\sigma \mathrm{Tr}\{\boldsymbol{\rho}_e \psi_{\sigma'}^\dagger(\mathbf{x}')\psi_\sigma^\dagger(\mathbf{x})\psi_\sigma(\mathbf{x})\psi_{\sigma'}(\mathbf{x}')\}$, respectively, where $\psi_\sigma(\mathbf{x})$ are operators corresponding to the particle spin components $\sigma = \uparrow, \downarrow$. Eq. (3) is valid for both fermionic and bosonic statistics, under the approximations detailed above.

The current correlations comprise two terms: a pair correlation term proportional to $G_e^{(2)}(\mathbf{x}',\mathbf{x})$, giving rise to coherent radiation when substituted into Eq. (2); and a term proportional to the probability density $G_e^{(1)}(\mathbf{x},\mathbf{x})$, contributing an incoherent radiation[39]. In this work, we focus on the case of a single particle, wherein $G_e^{(2)}(\mathbf{x}',\mathbf{x}) = 0$, and discuss the nature of quantum decoherence of the light it emits. A derivation of the effects of many-body quantum correlations ($G_e^{(2)}(\mathbf{x}',\mathbf{x}) \neq 0$) on the radiation will be reported in a future work.

**Cherenkov radiation**. CR is characterized by a directional, polarized, cone-shaped radiation pattern with opening semi-angle $\theta_c$ satisfying $\cos\theta_c = 1/\beta n(\omega)$, where $\beta = v/c$ is the speed of the particles normalized by the speed of light and $n(\omega)$ is the refractive index of the medium. We assume that the emission is detected with a far field detector located at a specific azimuthal angle on the cone's rim, providing broadband detection of all frequency components[63] (note that in certain practical situations, the entire emission ring (over all azimuthal angles) could be collected using special optics[64] – thereby increasing the signal level). Using the far-field expression for the dyadic Green tensor of a uniform dielectric medium[61], $\mathbf{G}(\mathbf{r},\mathbf{r}',\omega) = \frac{e^{iqr}}{4\pi r}(\mathbf{I} - \hat{\mathbf{r}}\hat{\mathbf{r}})e^{-i\mathbf{q}\cdot\mathbf{r}'}$, and assuming weak material dispersion, we find from Eqs. (2) and (3) that the radiation field projected on the detector is described by the

following frequency-domain quantum autocorrelation (see Supplementary Material section S3 for derivation)

$$\langle E^{(-)}(r,\omega')E^{(+)}(r,\omega)\rangle = \frac{U_0}{2\pi}\frac{e^{i(q_\omega-q_{\omega'})r}}{2n\epsilon_0 cr^2}\int d^3\mathbf{x}\, e^{i(\mathbf{q}_\omega-\mathbf{q}_{\omega'})\cdot\mathbf{x}} G_e^{(1)}(\mathbf{x},\mathbf{x})\ ,\quad (4)$$

where we denote $\mathbf{q}_\omega = \hat{\mathbf{r}}_c q_\omega$, with $\hat{\mathbf{r}}_c$ being the observation direction on the Cherenkov cone, $q_\omega = n(\omega)\omega/c$, and with $U_0 = \hbar\omega\alpha\beta\sin^2\theta_c$. Eq. (4) implies that for a single emitting particle, the first-order autocorrelation of CR is intimately related - through a Fourier transform - to the *probability density* of the particle wavefunction. The same conclusion – yet with more complex expressions – applies to all coherent CL processes, such as Smith-Purcell and transition radiation.

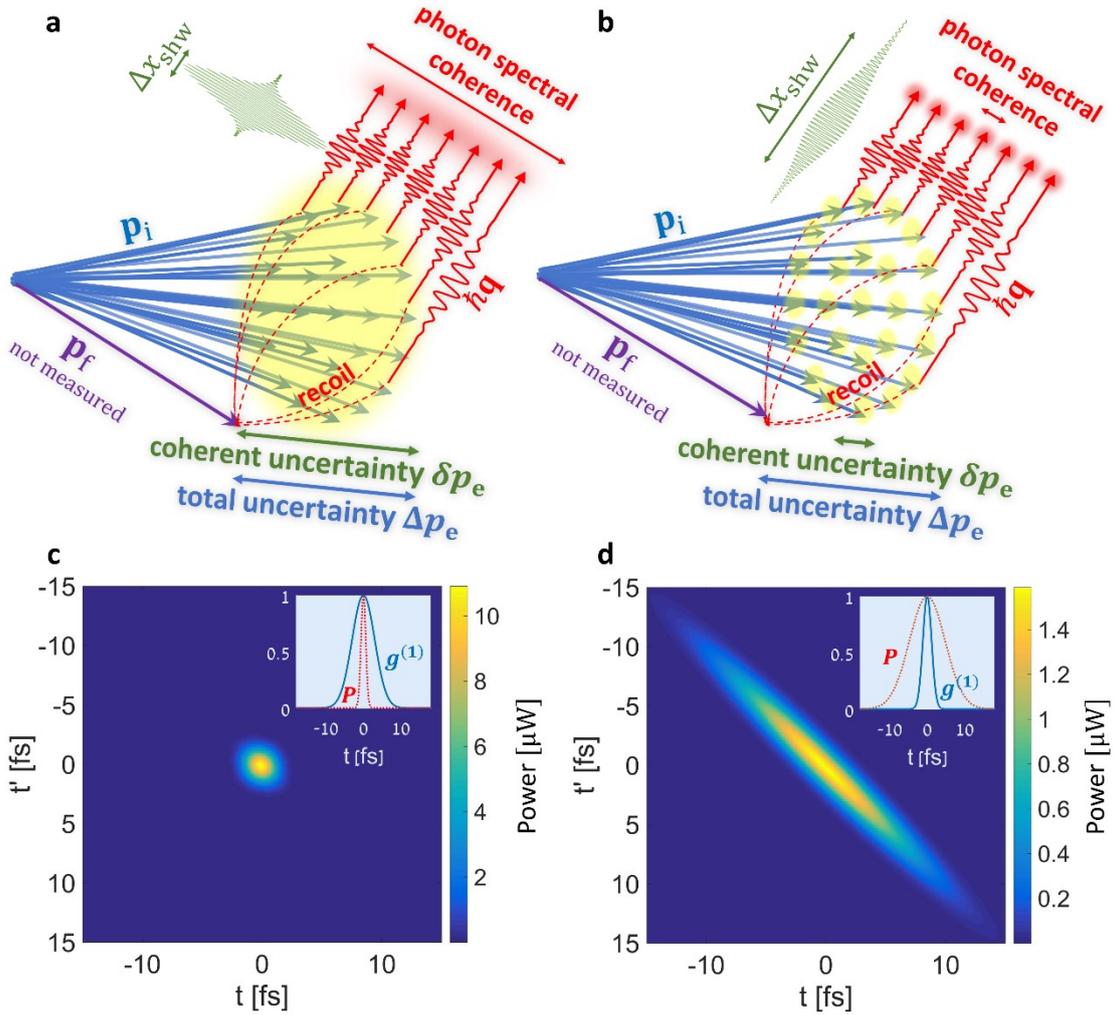

**Fig. 2: How can particle momentum uncertainty determine the interference of waves emitted by that particle?** (a) A quantum particle with a coherent momentum uncertainty $\delta p_e$ that equals its total momentum uncertainty $\Delta p_e$, displays a broad quantum coherence between its initial momenta $\mathbf{p}_i$ (yellow glow). When the particle transitions to any final momentum $\mathbf{p}_f$, the emitted wave inherits this initial coherence because of the "which path" interference between the initial particle states. Hence, different wavevector components of the wave are coherent (red glow). (b) A quantum particle in a mixture of momenta (total uncertainty $\Delta p_e$) with low coherent uncertainty $\delta p_e \ll \Delta p_e$ emits temporally incoherent waves. The limited interference inhibits the pulse formation, and its length exceeds the classical prediction. (c-d) The temporal field autocorrelations, $2r^2\epsilon_0 nc\langle E^{(-)}(t')E^{(+)}(t)\rangle$ (in µW), for 1 MeV electrons in silica in the visible range. The electrons are modelled as spherical Gaussian wavepackets with coherent energy uncertainty (a) $\Delta\varepsilon_e = 3.72$ eV (wavepacket radius ~50nm) and (b) $\Delta\varepsilon_e = 0.19$ eV (wavepacket radius ~1 µm). The diagonal ($t = t'$) indicates the temporal power envelope, $P(t)$, being transform limited in (a)

and incoherent in (b). Insets show a scaled comparison between $P(t)$ and the degree of first order coherence of the light, $g^{(1)}(\tau)$. For both (a) and (b), the classically expected shockwave FWHM is 1.4 fsec.

**Shockwaves from quantum particles: decoherence and a generalized uncertainty principle.**
The emission of classical shocks from a point charge[12], for which $\mathbf{j}(\mathbf{r},t) = e\mathbf{v}\delta(\mathbf{r} - \mathbf{v}t)$, is optically coherent over an arbitrarily wide spectral range only limited by the optical response of the medium. As such, the measured duration of the shockwave intensity envelope $|E(t)|^2$ is only limited by the material dispersion and/or the detection bandwidth, theoretically enabling shockwaves on the scale of femtoseconds and below[11,12]. The quantum description, however, incorporates the finite-sized single-particle wavefunction through Eq. (4). The incoherent emission from different points on the wavefunction (resulting from the delta-function term in Eq. (3)), is a manifestation of quantum decoherence of the emitted light, expected to inhibit interference visibility and stretch the shock duration.

Considering weakly dispersive media and wide detection bandwidths, the shockwave power envelope $P(t) = 2r^2\epsilon_0 nc\langle E^{(-)}(t)E^{(+)}(t)\rangle$ travelling at a group velocity $v_g$ is given by the equal-time temporal Fourier transform of Eq. (4). The probability cloud $G_e^{(1)}(\mathbf{x},\mathbf{x})$ is projected along the direction of observation $\hat{\mathbf{r}}_c$ on the Cherenkov cone. This relation implies that if the emitting particle wavefunction has a position uncertainty $\Delta x_e$ in the Cherenkov direction $\hat{\mathbf{r}}_c$, then the position uncertainty of the shockwave, $\Delta x_{\text{shw}}$, equals that of the emitting particle, i.e. $\Delta x_{\text{shw}} = \Delta x_e$. At this point, classical and quantum theories may seem to agree whenever $\Delta x_e$ becomes arbitrarily small. However, as quantum particles must obey the Heisenberg uncertainty principle[56] with respect to position and momentum, $\Delta x_e \Delta p_e \geq \hbar/2$, we obtain a generalized uncertainty principle for the shockwave

$$\Delta x_{\text{shw}} \Delta p_e \geq \frac{\hbar}{2}, \quad (5)$$

where an equality strictly holds for minimum uncertainty emitter states ($\Delta x_e \Delta p_e = \hbar/2$) and for broadband detection.

The simple yet surprising Eq. (5) demonstrates how the well-known classical result of wave interference can only be generated by a quantum particle that has a certain momentum uncertainty. This result also provides a fundamental quantum lower bound on the interference (the shockwave duration) that cannot be captured within a classical theory considering point particles, nor with a semiclassical theory treating the wavefunction as a coherent spread-out charge density[39,52] (see discussion below). As a concrete example, Fig. 2 shows how the momentum *coherence* $\delta p_e \leq \Delta p_e$ (or *coherent* momentum uncertainty) determines the lower bound on the shock duration. For example, particles in a mixed quantum state in momentum space, with low coherent uncertainty $\delta p_e \ll \Delta p_e$, emit temporally incoherent light and, consequently, a longer shockwave. These kinds of considerations also show why low frequency radiation (radio frequency, microwave, etc.) will generally be classical.

Experimentally, the decoherence effect best manifests itself in the temporal (or spectral) autocorrelations $\langle E^{(-)}(t)E^{(+)}(t')\rangle$ (or $\langle E^{(-)}(\omega)E^{(+)}(\omega')\rangle$), where the off diagonal ($t \neq t'$ or $\omega \neq \omega'$) terms relate to the coherence. Temporally coherent CR results in a transform-limited shockwave, as the classical theory suggests. However, the quantum corrections may alter the temporal behavior: coherent (incoherent) shockwaves exhibit $g^{(1)}(\tau)$ wider (narrower) than the pulse envelope. Fig. 2c-d demonstrates this behavior by simulating CR emission from

1 MeV electrons in silica for varying uncertainties. In this context, it is noteworthy to mention that classical and semiclassical theories predict that the emitted radiation is *always* perfectly coherent, both temporally and spectrally. Subsequently, it can be shown that these theories do not satisfy the quantum uncertainty principle Eq. (5) in practical experimental situations (e.g., in standard electron microscopes). An elaborate comparison between these theories can be found in the Supplementary Material, Section S7, and other works[37,39,52].

**Measuring the particle wavefunction dimensions using Cherenkov detectors.** Quantum optical measurement of the spectral autocorrelations may unveil information about the emitter wavefunction itself and provide an unprecedented analytical tool for particle identification. Eq. (4) provides a direct relation between the frequency-domain autocorrelation $\langle E^{(-)}(\omega')E^{(+)}(\omega)\rangle$ and the spatial Fourier transform (or structure factor) of the emitter probability density $G_e^{(1)}(\mathbf{x}, \mathbf{x})$. This structure factor is equivalent to a *momentum coherence* function of the particle, $\varrho_e(\mathbf{q}_\omega - \mathbf{q}_{\omega'}) = \int d^3\mathbf{k}\, \rho_e(\mathbf{k} + \mathbf{q}_{\omega'}, \mathbf{k} + \mathbf{q}_\omega)$, namely:

$$\langle E^{(-)}(\omega')E^{(+)}(\omega)\rangle \propto \int d^3\mathbf{x}\, e^{i(\mathbf{q}_\omega - \mathbf{q}_{\omega'})\cdot \mathbf{x}} G_e^{(1)}(\mathbf{x}, \mathbf{x}) = \varrho_e(\mathbf{q}_\omega - \mathbf{q}_{\omega'}), \qquad (6)$$

Eq. (6) implies that a spontaneously emitted photon is only as spectrally coherent as the emitting particle it originated from (see Fig. 2a-b). Spontaneous CR can, therefore, be used to map the structure of the emitter wavefunction and its momentum coherence by analyzing the correlations of emitted photons. Note that only the *spectral coherence* of the photons plays a role here, namely, the width of the off-diagonal part of the autocorrelations [see Fig. (3c)]. The diagonal part (optical power spectrum), $\langle E^{(-)}(\omega)E^{(+)}(\omega)\rangle$, is wavefunction independent – in accordance with previous theoretical[37] and experimental[39] studies.

The wavefunction dimensions can be estimated, for example, by assuming a spatial variance matrix for the emitter probability cloud given as $\mathbf{\Delta}_{ij}^2 = \text{Tr}\{\mathbf{r}_i \mathbf{r}_j \boldsymbol{\rho}_e\}$. The photons are collected at an observation direction $\hat{\mathbf{r}}_c$ on the Cherenkov cone, and the width of their spectral coherence $\Delta\omega$ is measured [see Fig. (3a-c)]. It then gives an estimate for the wavefunction dimensions $\mathbf{\Delta}$ along the observation direction (see Supplementary Material section S4 for the derivation)

$$\hat{\mathbf{r}}_c^T \mathbf{\Delta}^2 \hat{\mathbf{r}}_c = \frac{v_g^2}{\Delta\omega^2}, \qquad (7)$$

where $v_g$ denotes the shockwave group velocity. If the electron wavefunction is not spherical, we can further reconstruct the 3D $\mathbf{\Delta}$ by measuring the spectral coherence along different Cherenkov cones $\hat{\mathbf{r}}_c$ (which can be done by measurements of multiple particles with the same wavefunction moving through media of different refractive indices $n$, as done in threshold detection[19]). At least two such measurements are necessary to find both the longitudinal and transverse sizes of the wavefunction.

The quantum optical measurements necessary for the reconstruction of the photon density matrix in the frequency domain have been demonstrated experimentally for single photons[65–67]. Combining these quantum optical reconstruction techniques with Cherenkov detectors may allow for completely new and exciting capabilities. Currently available techniques for particle identification in Cherenkov detectors are limited to measuring velocity or mass[19,20]. Our proposed scheme further enables the measurement of the *wavefunction dimensions* and coherences of naturally occurring particles such as in cosmic radiation and beta decay[19], as well as the characterization of charged particle beams (for example in microscopy). Figs. 3d-e show an example for such measurement scheme for the case of 1 MeV electrons. This method can provide an alternative to matter wave holography (employed in electron microscopes[68] to measure the transverse wavefunction), which is currently unavailable for high-energy charged particles, such as muons, protons, kaons and pions. In contrast, the measurement we propose is relevant for these particles, and can be used as part of Cherenkov detectors, which also have the advantage of being a non-destructive measurement.

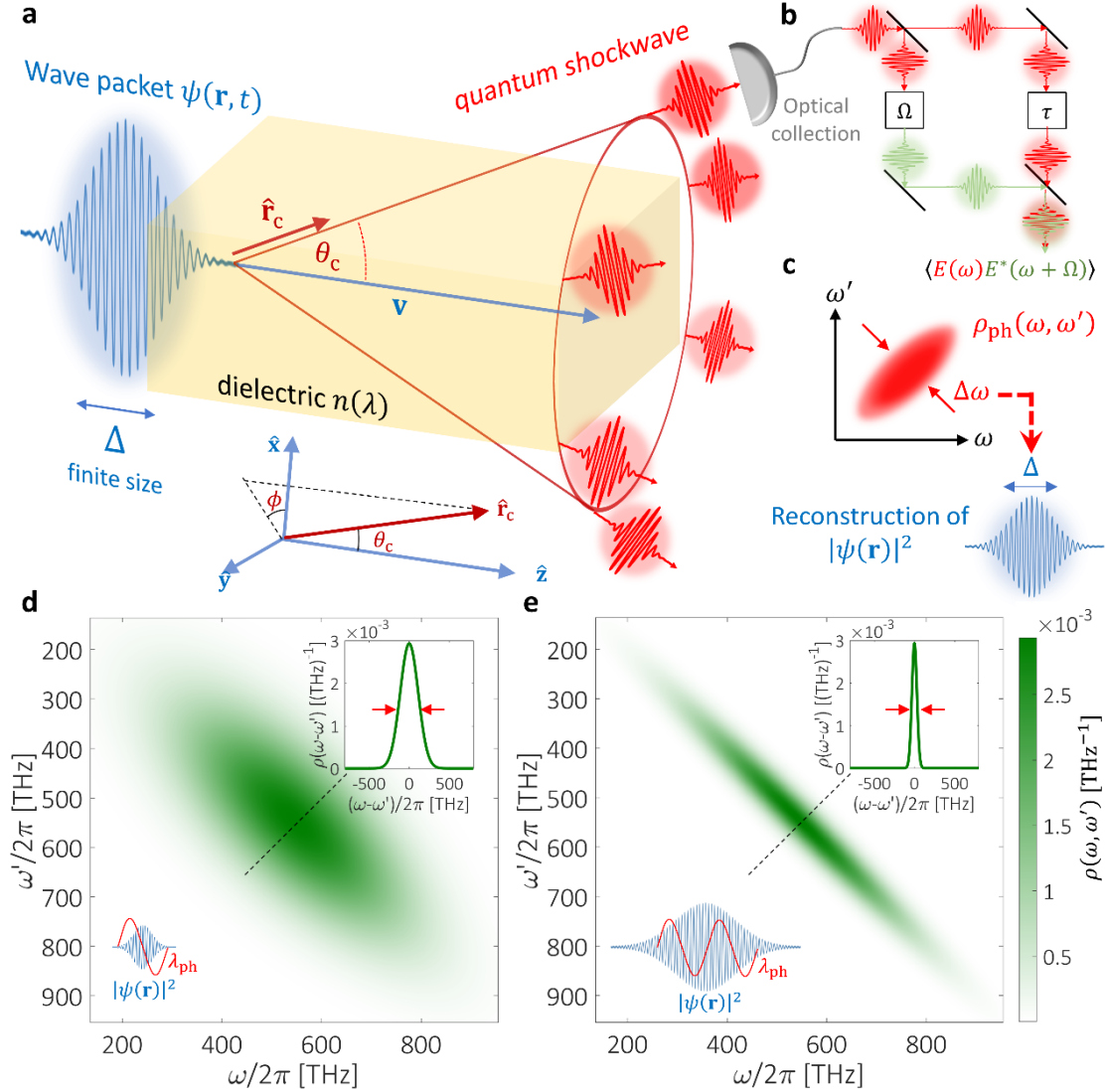

**Fig. 3: Quantum optical analysis of Cherenkov radiation – for measuring the emitter's wavefunction.** (a) A charged particle wavepacket $\psi(\mathbf{r}, t)$ of finite size $\Delta$ and carrier velocity $\mathbf{v_0}$ impinges on a Cherenkov detector with material dispersion $n(\omega)$. The particle spontaneously emits quantum shockwaves of light into a cone with opening half-angle $\theta_c(\omega) = \mathrm{acos}[1/\beta n(\omega)]$. Collection optics is situated along the cone in the direction $\hat{\mathbf{r}}_c$ in the far-field. (b) Detection scheme for measuring the spectral field autocorrelations $\langle E(\omega) E(\omega') \rangle$ utilizing an interference between spectrally/temporally sheared fields[65]. (c) The reconstructed photon density matrix determines the spatial probability distribution $|\psi(\mathbf{r})|^2$. (d-e) Simulation of particle wavefunction size reconstruction from the photon density matrix. A single 1 MeV electron ($\beta = 0.94$) in a silica Cherenkov detector (dispersion taken from Ref.[69]) emits Cherenkov radiation that is collected within the visible range ($\lambda = 400 - 700$ nm, centred at $\lambda_0 = 550$ nm). The electron wavefunction envelope is Gaussian and spherically symmetric, with position uncertainty of (d) $\Delta x_e = 254$ nm and (e) $\Delta x_e = 1016$ nm (bottom insets). In both (d) and (e), the measured photon density matrix,

$\rho_{\text{ph}}(\omega, \omega')$ is plotted. The wavefunction-independent diagonal $\rho_{\text{ph}}(\omega, \omega)$ that denotes the photodetection probability is the same for both cases. However, the off-diagonal spectral coherence $\rho_{\text{ph}}(\omega - \omega')$ is strongly dependent on the wavefunction. Measuring its width $\Delta\omega_{\text{coh}}$ (top insets) and employing the approximate Eq. (7) provides the estimates (d) $\Delta\tilde{x}_e = 290$nm and (e) $\Delta\tilde{x}_e = 1006$nm.

Beyond the capability of reconstructing the wavefunction size, our technique can be used to detect the signature of non-Gaussian wavepackets (in energy-time space), such as coherent electron energy combs produced in photon-induced near-field electron microscopy (PINEM)[34,70] (see Fig. 4). In PINEM, a free electron traverses a near-field optical structure and interacts with a coincident laser pulse. As a result, the electron wavefunction is modulated, and given by a coherent superposition of energy levels. Following free-space propagation, the electron wavefunction takes the form of a pulse train[34]. Notice how the interference fringes due to the shaped wavefunction appear only in the photon spectral autocorrelations (off-diagonal), and not the radiation spectrum (diagonal).

**Experimental considerations.** Here, we briefly discuss some important considerations for realizing our predictions in an experiment. For the analysis discussed in the previous section, the electron's coherent interaction length $L_{\text{int}}$ must be in the range $\lambda/n \ll L_{\text{int}} \ll (n/\Delta n)(\lambda/\Delta\lambda)\beta\lambda$, where $\Delta n = n - n_g$ is the difference between the refractive and group indices of the material, and $\Delta\lambda$ denotes the wavelength band collected by the detection system (see Supplementary Material, Section S5). For standard materials and optical wavelengths, $L_{\text{int}}$ is of the order of a few microns.

For Cherenkov radiation in bulk media, other scattering processes with mean free paths smaller than $L_{\text{int}}$ can readily broaden the particle spectrum $\rho_e(\mathbf{k}, \mathbf{k})$. In the Supplementary Material Section S6, we show that for a general uniform medium of optical response function[71] Im $\mathbf{G}(\mathbf{q}, \omega)$, the momentum coherence function $\varrho_e(\mathbf{q}_\omega - \mathbf{q}_{\omega'})$ of Eq. (6) remains unchanged. As the latter quantity is the one responsible for the Cherenkov autocorrelation through Eq. (6), we expect the signature of the wavefunction to persist. In electron microscopy, one can avoid these scattering processes by employing an aloof beam geometry having electrons that propagate in vacuum near an optical structure, such as in Smith-Purcell experiments[72] or in emission of Cherenkov photons near dielectric boundaries[73].

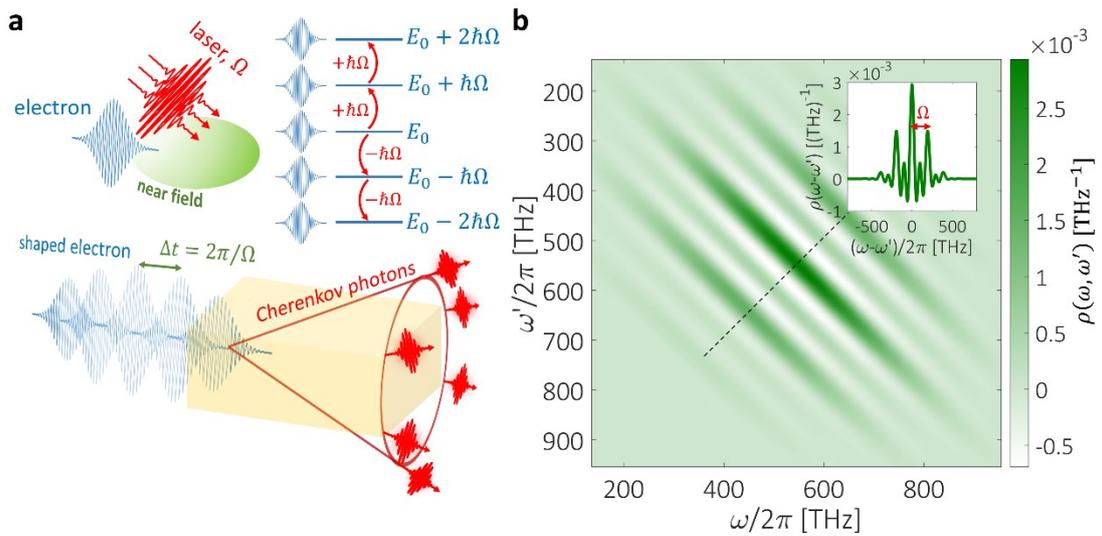

**Fig. 4 Quantum optical analysis of Cherenkov radiation – emitted from a laser-driven electron wavefunction.** (a) a free electron wavefunction is shaped by the interaction with a strong laser field of frequency $\Omega$ (here $\Omega = 2\pi \times 200$ THz), as done in photon induced near-field electron microscopy[70]. The result is a coherent electron

energy ladder, manifested as a temporal pulse train. (b) Cherenkov photon autocorrelations reveal the electron wavefunction spectral interference pattern, matching the laser frequency. The measurement scheme is the same as in Fig. 3a-c.

**Discussion**

In this paper, we investigated light emission by free charged particles from a quantum-optical viewpoint, by employing a fully-quantum formalism of light–matter interaction. Our conclusions take into account the experimental situation that the emitting particle itself is not measured. Importantly in this situation, recent studies show that the particle wavefunction has no influence on the emitted spectrum. We complement this realization by showing that quantum optical measurables such as the emitted pulse duration and optical autocorrelations are all strongly influenced by the particle wavefunction. Moreover, all the quantum features of the particle such as coherence, uncertainty and correlations embedded in the emitter wavefunction play an important role in determining the properties of the emitted light.

As an example, we considered the Cherenkov effect, and its characteristic optical 'shockwave', envisioned classically for almost a century as a coherent, transform-limited pulse of light. Instead, we found it is fundamentally limited by the particle quantum uncertainty, satisfying a generalized uncertainty principle. The smaller the coherent momentum uncertainty of the particle is, the longer (and less coherent) the shockwave becomes. We further showed how this uncertainty relation can be harnessed to unveil information about the particle wavefunction, allowing unprecedented capabilities for particle detection. For example, Cherenkov detectors together with a quantum-optical measurement of the emitted light can be employed to reconstruct the particle wavefunction size, shape, coherence and quantum correlations.

Our findings can be employed to resolve an important fundamental question of practical importance: what part of the energy uncertainty of a free electron is coherent, and what part is incoherent? This property can be measured from the radiation autocorrelations and spectrum. With the advent of laser-driven electron sources, for example in ultrafast electron microscopes[32–34,74–77], the particle coherent energy uncertainty is believed to be dictated by the laser linewidth[33,78], e.g. spanning tens of meV for excitations with femtosecond lasers. With the ability to coherently control the spatial electron wavefunction[30,79–82], the transverse momentum uncertainty can be further lowered. Such conditions allow for the predictions of our work to be tested experimentally under controllable settings.

Considering the outlook for using free electrons as quantum probes[18], our work paves the way towards quantum measurement of free electrons and other charged particles based on spontaneous emission. Our results may readily be generalized to other physical mechanisms of wave emission, for example analogues of the Cherenkov effect[83–86] outside of electrodynamics, as in Bose-Einstein condensates. Similar effects can be explored with any photonic quasiparticle[87], and even with sound waves, and phonon waves in solids[88], which all have the same underlying quantum nature and must have exact analogous phenomena.

Another intriguing question is the effect of many-body correlations (as manifested by the second term in Eq. 2) on such radiation phenomena, giving rise to yet unexplored quantum super- and subradiance regimes of coherent CL. These arise from coherent interference of multi-particle wavefunctions (resembling the Hong-Ou-Mandel effect[89]), which will be discussed in forthcoming work.